\documentclass[12pt]{article}
\pdfoutput=1

\usepackage{amsmath,amssymb,amsfonts}
\usepackage{graphicx}
\usepackage[pdftitle={effective fluids},pdfauthor={Guillermo Ballesteros & Brando Bellazzini},colorlinks=true,citecolor=black,linkcolor=black,urlcolor=blue]{hyperref}
\usepackage{color}
\usepackage{slashed}
\usepackage{framed}				
\usepackage{verbatim}
\usepackage{cite}

\newcommand{\II}{\mbox{${\mathbb I}$}}

\newcommand{\ch}{\mathcal H}
\newcommand{\eq}[1]{(\ref{#1})}

\numberwithin{equation}{section}
\definecolor{MyRed}{rgb}{0.9,0.12,0.1}
\definecolor{MyBlue}{rgb}{0.1,0.12,0.9}

\newcommand{\arXiv}[2]{\href{http://arxiv.org/pdf/#1}{{\tt #2/#1}}}
\newcommand{\arXivold}[1]{\href{http://arxiv.org/pdf/#1}{{\tt #1}}}

\usepackage[english]{babel}

\begin{document}
\begin{titlepage}


\begin{center}
\begin{huge}
\textbf{Effective perfect fluids in cosmology}
\end{huge}
\end{center}

\vskip 1.0 cm

\begin{center}
\begin{large}{\bf Guillermo Ballesteros}$^{1,2,4}$ and {\bf Brando Bellazzini}$^{2,3}$\end{large}
\end{center}
\vskip 0.5 cm

\begin{center}
$^{1}${\it Museo Storico della Fisica e Centro Studi e Ricerche Enrico Fermi,\\ Piazza del Viminale 1, I-00184 Rome, Italy}\\
\vskip 0.2cm

$^{2}$ {\it  Dipartimento di Fisica, Universit\`a di Padova and INFN, Sezione di Padova,\\Via Marzolo 8, I-35131 Padova, Italy} \\
\vskip 0.2cm

$^{3}$ {\it   SISSA,  Via Bonomea 265, I-34136 Trieste, Italy} 
\vskip 0.2cm

$^{4}$ {\it Universit\'{e} de Gen\`eve, Department of Theoretical Physics and \\Center for Astroparticle Physics (CAP), \\24 quai E. Ansermet, CH-1211 Geneva 4, Switzerland}

\vskip 0.74cm

{\tt \small 
 \href{mailto:guillermo.ballesteros@unige.ch}{\color{black} guillermo.ballesteros@unige.ch}, 
 \href{mailto:brando.bellazzini@pd.infn.it}{\color{black} brando.bellazzini@pd.infn.it}}

\end{center}

\vskip 1.45cm

\begin{abstract}

We describe the cosmological dynamics of perfect fluids within the framework of effective field theories.  The effective action is a derivative expansion whose terms are selected by the symmetry requirements on the relevant long-distance degrees of freedom, which are identified with comoving coordinates. The perfect fluid is defined by requiring invariance of the action under internal volume-preserving diffeomorphisms and general covariance. At lowest order in derivatives, the dynamics is encoded in a single function of the entropy density that characterizes the properties of the fluid, such as the equation of state and the speed of sound. This framework allows a neat simultaneous description of fluid and metric perturbations. Longitudinal fluid perturbations are closely related to the adiabatic modes, while the transverse modes mix with vector metric perturbations as a consequence of vorticity conservation.
This formalism features a large flexibility which can be of practical use for higher order perturbation theory and cosmological parameter estimation. 

\end{abstract}

\end{titlepage}

%
%


\section{Introduction}

The success of effective field theories for studying physical systems comes from the fact that they allow to capture in a single picture the universal long distance (low energy) properties of models that are instead intrinsically different at much shorter scales.  Since cosmological problems are often characterized by well separated scales, the language of effective field theories provides a powerful tool for the study of cosmological evolution, in particular inflationary and dark energy dynamics.
Most of the works on cosmological evolution that are based on effective theories can be broadly classified in two different categories,  depending on whether they aim to describe the full (effective) action \cite{Weinberg:2008hq,Burgess:2009ea,Barbon:2009ya,Park:2010cw,Bloomfield:2011np,Burgess:2012dz} or, conversely, focus on the dynamics of the perturbations around some background \cite{Creminelli:2006xe,Matarrese:2007wc, Cheung:2007st,Cheung:2007sv,Creminelli:2008wc,Senatore:2009cf,Bartolo:2010bj,Baumann:2010rc,Bartolo:2010di,Senatore:2010wk,Bartolo:2010im,Creminelli:2010qf,
Creminelli:2011rh,Pietroni:2011iz,LopezNacir:2011kk,Senatore:2012nq,Pimentel:2012tw,Senatore:2012wy,Achucarro:2012yr,Carrasco:2012cv,Hertzberg:2012qn,Nacir:2012rm,Gubitosi:2012hu,Gwyn:2012mw}. The formalism we develop in this work pertains to the former class but, as we will see,  gives a general and straightforward effective expansion for cosmological fluid perturbations on a Friedmann-Lema\^{i}tre-Robertson-Walker (FLRW) metric (and may also be extended to other backgrounds). 

There has recently been a renewed interest in the understanding of fluid dynamics from an effective field theory point of view \cite{Dubovsky:2005xd,Endlich:2010hf,Dubovsky:2011sj,Nicolis:2011cs,Nicolis:2011ey,Dubovsky:2011sk,Torrieri:2011ne,Hoyos:2012dh}. 
This general approach is based on the identification of the relevant long-distance degrees of freedom and their symmetries, possibly including spacetime symmetries such as Galileo, Poincar\'e,  or diffeomorphism invariance. The dynamics is organized in a systematic derivative expansion that makes the theory predictive at low energies.  
This formalism (or variations of it) has already been applied to describe some aspects of perfect fluids and superfluids \cite{Nicolis:2011cs,Nicolis:2011ey,Dubovsky:2011sk,Hoyos:2012dh}. However, although fluids are ubiquitous in cosmology, the cosmological applications of this framework have been, to the best of our knowledge, largely neglected\footnote{See however \cite{Blas:2012vn}  for an application of this formalism to a Lorentz violating dark matter model.}.  In this work we describe the basic set-up for perfect fluids, adopting the general principles of effective field theory to describe long-distance relativistic dynamics in cosmology and their interplay with metric perturbations.  

Neglecting chemical potentials, we consider perfect (dissipationless) fluids that carry no conserved charge.  We thus focus on \textit{minimal} fluids that involve only three degrees of freedom associated with the position in space of a fluid element. These degrees of freedom can be identified with comoving coordinates and, as we will see, the symmetries of the fluid determine the operators that appear in the action. In particular, perfect fluids are invariant under internal spatial diffeomorphisms that preserve the volume. As a result, the relativistic dynamics is fully described at lowest order in derivatives by a single operator. By construction, these fluids support both longitudinal (compressional) and transverse (vortices) excitations, each with its own dynamics.  We discuss perturbation theory for a fluid with an arbitrary equation of  state, and describe the coupling to scalar, vector and tensor metric perturbations. Vector metric perturbations mix with vortices, while adiabatic perturbations correspond to the compressional modes of the fluid. Our formalism can cover and extends other frameworks (e.g. $P(X)$ theories, where the dynamics of fluid vortices is instead absent).

Following the formulation developed in  \cite{Dubovsky:2005xd, Endlich:2010hf, Dubovsky:2011sj, Nicolis:2011cs}, which can be connected to earlier works \cite{Leutwyler:1996er,Andersson:2006nr,Brown:1992kc,Comer:1993sp,Comer:1994tw,bookcarter},  we extend the effective theory of perfect fluids to generic metric backgrounds in Section~\ref{effective}. We then focus on the FLRW case in Section~\ref{flrw}, and move on to discuss matter and metric  perturbations in Sections~\ref{perturb1} and \ref{metricp}. Adiabatic perturbations are introduced in Section~\ref{genad}; and the conclusions and future work directions are presented in \ref{conclusions}. In the Appendix \ref{appa} we review perfect fluids in the context of relativistic hydrodynamics. In the Appendix~\ref{appEulerLagrange} we comment on the relation between Eulerian and Lagrangian formulations for fluids, which we both use through the text. We discuss vorticity conservation using ADM variables \cite{Arnowitt:1962hi} in Appendix \ref{vorticitymetric}.

\section{The effective theory of perfect fluids} \label{effective}

In this section we will closely follow the formulation of perfect fluids of \cite{Dubovsky:2005xd, Endlich:2010hf, Dubovsky:2011sj, Nicolis:2011cs} which recasts in the modern language of the effective field theories some earlier results of the pull-back approach \cite{Andersson:2006nr,Brown:1992kc,Comer:1993sp,Comer:1994tw,bookcarter} which is based on the action formalism pioneered in \cite{Taub:1954zz,Carter:carter}. 

\begin{figure}[tb]
\centering
\includegraphics[trim = 0 435 265 100 clip, width=12cm]{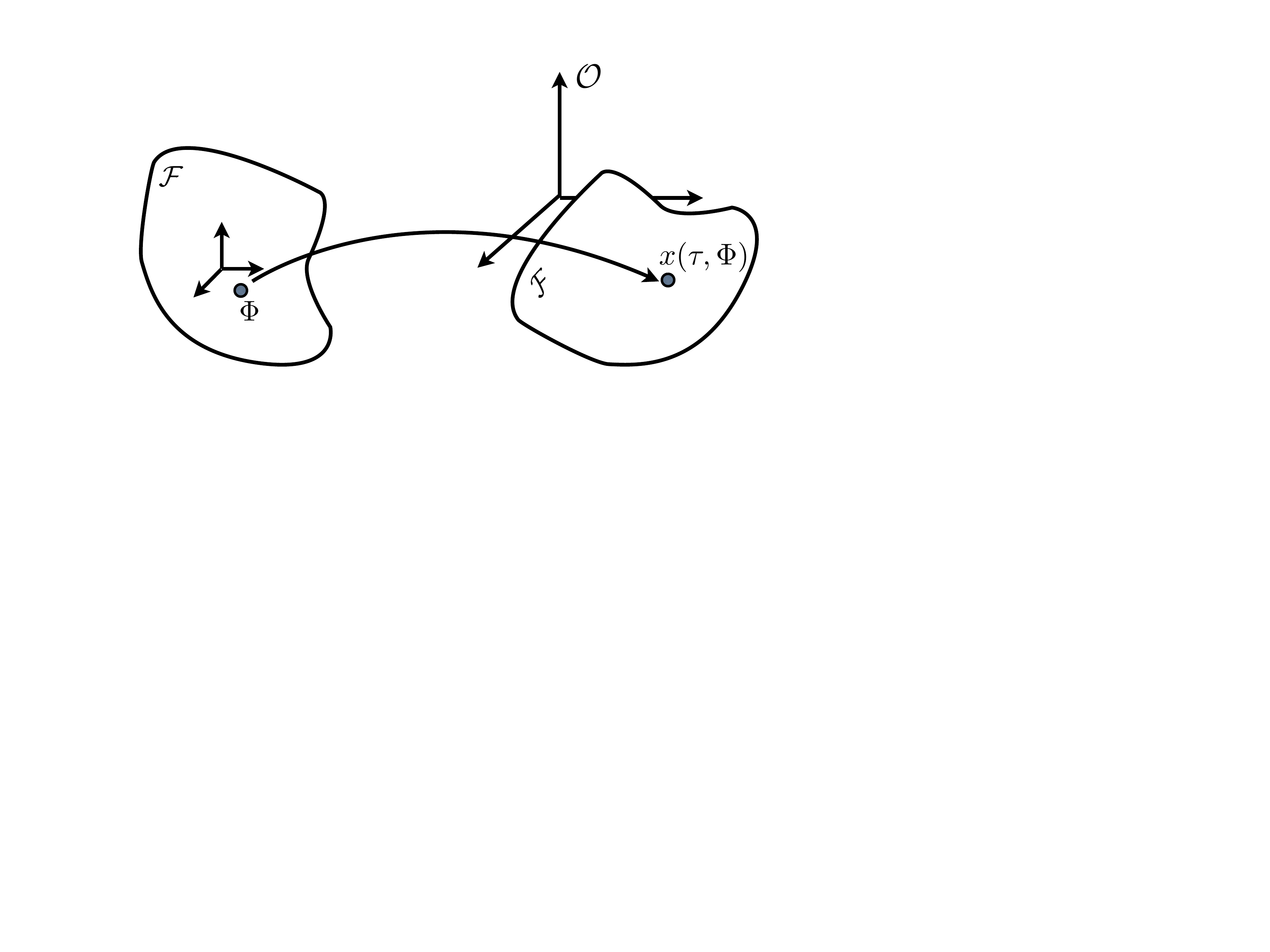}
\caption{\label{isomor} {\small The map between $\Phi$ and $x$ coordinates is depicted. At any given (conformal) time $\tau$\,, a fluid element labelled by $\Phi$ occupies a position given by $x(\tau,\Phi)$.  If the inverse function is considered, any spacetime point $(\tau,x)$ is mapped to a fluid element $\Phi$. In this picture, the $\Phi$ coordinates are scalar fields of spacetime.}}
\end{figure}

A perfect fluid is described by a functional of three spacetime scalar functions $\Phi^{a}\in \mathbb{R}$ with $a\in\{1,2,3\}$\,, that define, at any time, an isomorphism between the three-dimensional coordinate space of an observer $\mathcal{O}$ and a continuum of points $\mathcal{F}$ (the fluid) \cite{Dubovsky:2005xd,Andersson:2006nr}.  These functions $\Phi^a$ label a generic fluid element\footnote{We are focusing on \textit{mechanical} fluids that  at each point in space are fully described by three degrees of freedom. See Appendix \ref{appEulerLagrange} for more details and the relation between these fluids and the continuum limit of  relativistic uncharged point-like particles.} in $\mathcal{F}$, whereas the map $\Phi^a\longrightarrow x^i(\tau,\Phi)$ gives the position in real space of the fluid element $\Phi$ at a given time $\tau$.
In other words, $x^i=x^i(\tau,\Phi)$ is the trajectory of the fluid element $\Phi$. This corresponds to a Lagrangian description of the dynamics, see Figure \ref{isomor}.  Vice versa, the inverse map $x^i\longrightarrow \Phi^a(\tau,x)$ returns the fluid element $\Phi$ that is sitting in $x$ at the time $\tau$, providing an Eulerian description of the dynamics.  In a stationary background state, the isomorphism can be chosen such that $\Phi^a(\tau,x^i)=x^a$.

 Since both $\mathcal{O}$ and $\mathcal{F}$ are isomorphic to $\mathbb{R}^3$, there is always a change of spacetime coordinates $\mathcal{O}\rightarrow\mathcal{\tilde O}$ such that  $x^i\mapsto \tilde x^i=\Phi^i(\tau,x^j)$ so that $\Phi^a$ can be naturally interpreted as the comoving coordinates of the fluid. 
 This means that their variation along the fluid four-velocity $u^\mu=dx^\mu/d\eta$ (being $\eta$ the proper time) is zero
\begin{equation}
\label{fluidvelocity}
u^\mu \partial_\mu \Phi^a(x,\tau)=0\,,\qquad u^\mu u_\mu=-1\,.
\end{equation}
As we will see next, these two conditions characterize completely the fluid four-velocity, once the symmetry properties of the system have been chosen. Indeed, the whole structure of the effective theory is fully determined by the symmetries of the fluid.

Since the fluid must be homogeneous and isotropic, the internal coordinates have to satisfy the symmetries
\begin{align}
\label{homogeneity}
\Phi^a \longrightarrow &\Phi^a+c^a\\
\label{isotropy}
\Phi^a \longrightarrow & R^a_b\Phi^b\,,\qquad R\in SO(3)
\end{align}
where $c^a$  and the matrix of elements $R^a_b$ are constant in space and time. In a homogeneous and isotropic model of the universe, even though the background solution $\Phi^a(x^i,t)=x^a$, which represents the ground state of the system, spontaneously breaks spatial translations and rotations, the diagonal combination of internal (acting on $\Phi$) and space (acting on $x$) symmetries is left unbroken \cite{Dubovsky:2005xd}. These diagonal symmetries ensure that the perturbations (or in other words, the excitations of the fluid) $\pi^a=\Phi^a-x^a$ propagate in a homogeneous and isotropic background.

 In addition, we demand invariance under volume preserving spatial diffeomorphisms
 \begin{equation}
 \label{diffs}
 \Phi^a\longrightarrow  f^a(\Phi)\,, \qquad  \det\left(\frac{\partial f^a}{\partial \Phi^b}\right)=1
 \end{equation}
that distinguish a perfect fluid from a gel (or jelly), which is a homogeneous and isotropic solid \cite{Dubovsky:2005xd}. It is clear that the spatial volume preserving diffeomorphisms \eq{diffs} include the symmetry transformations \eq{homogeneity} and \eq{isotropy}, but we have highlighted the latter two because of their clear geometrical meaning. 

The low energy effective theory for a perfect fluid is given by a  Lagrangian $\mathcal{L}$ organized in a derivative expansion. Since the dynamics is invariant under spacetime diffeomorphisms and the transformations (\ref{homogeneity}), (\ref{isotropy}) and (\ref{diffs}),  the lowest order Lagrangian must be a  function of the determinant of the matrix $B$, whose elements are given by \cite{Leutwyler:1996er,Dubovsky:2005xd}
\begin{align} \label{bmatrix}
B^{ab}\equiv g^{\mu\nu}\partial_\mu \Phi^a \partial_\nu \Phi^{b}\,.
\end{align} 
The effective action that describes the low energy dynamics of the fluid is then
\begin{align} \label{action}
S_m[\Phi]=\int d^4x \sqrt{-g}\,\mathcal{L}_m 
\end{align}
where the Lagrangian (density) is a function of the determinant of $B$
\begin{align}
\label{lagrangian}
\mathcal{L}_m =& F(b) \\
\label{smallb}
b\equiv & \sqrt{\det B}
\end{align}
that encodes the long distance properties of the fluid.  We are implicitly assuming that there are no extra symmetries that could forbid any possible Lagrangian $F(b)$. Otherwise, the low energy Lagrangian would start at the next order in derivatives: $g(b)u^\mu \partial_\mu b$, where $g(b)$ is an arbitrary function of $b$ \cite{Dubovsky:2011sj}. However,  this term can be recast into a higher derivative term by a field redefinition \cite{Dubovsky:2011sj}.  In the following we will consider only the lowest order Lagrangian (\ref{lagrangian}). 
 
The equations of motion that come from (\ref{lagrangian}) are 
\begin{equation} \label{eom}
\partial_\mu\left[\sqrt{-g}\, g^{\mu\nu}\, b\ F_b\,\,(B^{-1})^{cd}\partial_\nu\Phi^d\right]=0
\end{equation}
where $F_b$ denotes the derivative of $F(b)$ with respect to $b$. 
The gravitational energy-momentum tensor of the system is 
\begin{equation} \label{emt}
T_{\mu\nu}=-\frac{2}{\sqrt{-g}}\frac{\delta S_m}{\delta g^{\mu\nu}}=g_{\mu\nu}F-b\,F_b  (B^{-1})^{cd}\partial_\mu \Phi^c \partial_\nu \Phi^d\,.
\end{equation}
This corresponds to the energy-momentum tensor of a perfect fluid
\begin{equation} \label{pf}
T_{\mu\nu}=(\rho+p)u_\mu u_\nu+p g_{\mu\nu}
\end{equation}
whose components can be easily identified. The conditions \eq{fluidvelocity} determine the fluid four-velocity, which is \cite{Dubovsky:2005xd}
\begin{align}
u^\mu=-\frac{1}{6b\,{\sqrt{-g}}}\epsilon^{\mu\alpha\beta\gamma}\epsilon_{abc}\partial_\alpha\Phi^a\partial_\beta\Phi^b\partial_\gamma\Phi^c\,.
\end{align}
Contracting $T^{\mu\nu}$ with $u^\mu u^\nu$ (and recalling that $u^2=-1$) one gets, in agreement with early works on the action formalism for perfect fluids \cite{Taub:1954zz,Carter:carter}, that
\begin{equation} \label{rho}
\rho=-F
\end{equation}
and therefore the fluid is not only perfect but also {\it isentropic} \cite{Brown:1992kc}.
For a physical interpretation of $\mathcal{L}_{m}=-\rho$ see Appendix \ref{appEulerLagrange}.
Finally, from the trace $T_\mu^\mu$ we also obtain the pressure
\begin{equation} \label{trace}
\rho+p=-b\,F_b\,.
\end{equation}
The fluid is thus {\it barotropic} because the pressure depends only on the energy density given that $p$ and $\rho$ are both functions of  $b$ only.

We also find that
\begin{equation}
h_{\mu\nu}\equiv (B^{-1})^{ab}\partial_\mu \Phi^a \partial_\nu \Phi^b=g_{\mu\nu}+ u_\mu u_\nu
\end{equation}
is the standard projector on hypersurfaces orthogonal to the four-velocity of the fluid.

Notice that for dust (pressureless matter) $p=0$ and so  $F\propto b$, while for radiation  $\rho=3p$ and $F\propto b^{4/3}$. In general, the relation between the background energy density and pressure is, as usual, given by the equation of state, which is defined as
\begin{align}
{\bar p}=w \bar{\rho} 
\end{align}
and therefore using \eq{rho} and \eq{trace} this gives
\begin{align} \label{wminus1}
w=-1+\bar{b}\frac{\bar F_b}{\bar F}=-1+\overline{\frac{d \log F}{d \log b}}\,.
\end{align}
The current\footnote{We fix the normalization of $\mathcal{J}^\mu$ such that the temperature in (\ref{temp}) matches the standard thermodynamics normalization, $T=(\partial\rho/\partial s)$.}
\begin{align}
\mathcal{J}^\mu=-b\, u^\mu
\end{align}
is covariantly and identically (i.e. off-shell) conserved 
\begin{align}
 \mathcal{J}^\mu_{\,\,\,;\,\mu}=0
\end{align}
and it is identified with the entropy current \cite{Torrieri:2011ne,Dubovsky:2011sj}. The comoving entropy density $\mathcal{J}^\mu u_\mu=s$ becomes
\begin{align}
\label{helm}
s=b=\frac{\rho+p}{T}
\end{align}  where $T$ is the fluid temperature, given by
\begin{align}
\label{temp}
T=-F_b\,.
\end{align}
Starting from the entropy current $\mathcal{J}^\mu$ it is possible to construct  infinitely many other currents $\mathcal{K}^\mu_{(f)}=f(\Phi)\mathcal{J}^\mu$ \cite{Comer:1993sp}  which are identically conserved, $\mathcal{K}^\mu_{(f)\,;\,\mu}=0$, because $\Phi$ are comoving coordinates satisfying the equation (\ref{fluidvelocity}). These currents do not define an independent flow since they are all \textit{aligned} with the entropy current (and thus the fluid four-velocity). Therefore, the associated charge transfer occurs only along the direction of mechanical fluid displacement.  To each of these currents $\mathcal{K}^\mu_{(f)}$ one can associate a chemical potential $\mu_f$ generalizing the equation (\ref{helm}) to $\rho+p=Ts+\mu_f n_f$, including the contribution from the comoving charge density $n_f$.  In this work  we focus on fluids that carry none of these comoving charges, except for the entropy; so all chemical potentials vanish whereas $T\neq0$. While the approximation of vanishing chemical potentials is a good one  for many cosmological applications, finite charge densities can be easily incorporated by allowing non-vanishing chemical potentials \cite{Comer:1993sp,Dubovsky:2011sj}. 

The invariance under spatial diffeomorphisms (\ref{diffs}) gives rise, via Noether's theorem,  to another set of infinite  (on-shell) conserved currents \cite{Dubovsky:2005xd}
\begin{equation}
\mathcal{J}^\mu_{(\varepsilon)}=-b F_b\, (B^{-1})^{cd} \partial^\mu\Phi^d \varepsilon^c(\Phi)\qquad   \mathcal{J}^\mu_{(\varepsilon)\,\,\,;\,\mu}=0
\end{equation}
where $\varepsilon^a(\Phi)$ is an arbitrary transverse function of $\Phi$
\begin{equation}
\label{epsilondiff}
\partial_a \varepsilon^a(\Phi)=\frac{\partial\varepsilon^a}{\partial\Phi^a}=0\,.
\end{equation}
This constraint, generically solved by $\varepsilon^a(\Phi)=\epsilon_{abc}\partial_b f_c(\Phi)$, ensures that $\Phi^a\rightarrow \Phi^a+\varepsilon^a(\Phi)$ is an infinitesimal volume preserving diffeomorphism.
In the Appendix~\ref{vorticitymetric} we explicitly construct the associated conserved charges and comment on their relation with the vorticity conservation.

\section{Effective perfect fluids in FLRW}
\label{flrw}

We have seen that the perfect fluid form of the energy-momentum tensor \eq{emt} comes from imposing the conditions of homogeneity, isotropy and invariance of the action under volume preserving diffeomorphisms of the internal coordinates. In cosmology, the first two assumptions directly lead to the FLRW metric. In fact, the FLRW metric is a purely geometric consequence of requiring that the universe appears homogeneous and isotropic to free falling observers \cite{Weinberg:100595,Weinberg:1102255}. For simplicity we assume from now on that the background metric of the universe is of flat FLRW type and work with conformal time $\tau$,
\begin{align} \label{unpfrw}
ds^2=-a^2(\tau)(d\tau^2+\delta_{ij}dx^idx^j)\,.
\end{align}
The usual equations of motion that govern the background cosmology in the metric \eq{unpfrw} simply read:
\begin{align} \label{fried}
\ch^2=\frac{8\pi G}{3}a^2\bar\rho_T\qquad 
\dot\ch=-\frac{4\pi G}{3}a^2(\bar\rho_T+3\bar p_T)
\end{align}
where $\ch=\dot a/a$\,. Newton's gravitational constant is denoted by $G$ and the quantities $\bar\rho_T$ and $\bar p_T$ represent the total background energy density and pressure if several fluids are present. Let us point out that if there are several fluids that only interact among themselves through gravity, each of them satisfies the following Friedmann equation for the background 
\begin{align} \label{cons}
\dot{\bar\rho}_\alpha+3(1+w_\alpha)\ch\bar\rho_\alpha=0\,.
\end{align}

From the equations \eq{fried} and the expressions \eq{rho} and \eq{trace} one sees that all self-accelerating solutions (which require $w_T <-1/3$) have to satisfy the condition 
\begin{align}
3\bar{b} \bar F_{Tb}<2\bar F_T
\end{align}
where $F_T$ is the function \eq{lagrangian} that describes the total energy density of the fluid admixture. Notice also that \eq{wminus1} implies that a perfect fluid will have an equation of state that is close to that of a cosmological constant ($w\simeq -1$) if $|d \log F/d \log b|\ll 1$.

Given the FLRW metric  (\ref{unpfrw}), the matrix \eq{bmatrix} reads 
\begin{equation}
 \label{bmatrixfrw}
 B=\frac{1}{a^2}(\partial\Phi)^T \left(\II- v\otimes v \right)\partial \Phi
\end{equation}
 where $\II$ is the identity matrix, $v\otimes v$ is a matrix of elements $v^i v^j$, and we have defined
\begin{align} 
\label{velo}
(\partial\Phi)_i^a\equiv\partial_i \Phi^a\,,\qquad v \equiv-(\partial\Phi^T)^{-1}\dot\Phi
\end{align} 
where the dots denote derivatives with respect to conformal time $\tau$. 
Recalling that the total time derivative of $\Phi$  vanishes, i.e.
\begin{align} \label{vsolver}
\frac{d\Phi^a}{d\tau}=\frac{\partial\Phi^a}{\partial \tau}+\frac{\partial \Phi^a}{\partial x^j}v^j=0
\end{align}
one recognizes $v$ in \eq{velo} as the (Lagrangian) velocity
\begin{align}
v^i=\dot x^i
\end{align}
In the language of fluid dynamics, the equation (\ref{vsolver}) means that the material (or convective) derivative of $\Phi$ is zero, which is nothing else than the statement that the label of a fluid element does not change along its trajectory, accordingly with the interpretation of $\Phi$ as the comoving coordinates. 
This is equivalent to the first of the conditions \eq{fluidvelocity}. In fact, from the definition of the four-velocity, $u^\mu=dx^\mu/d\eta$, we get, consistently with equation \eq{vsolver}, 
\begin{align}
 \label{gammavel}
u^0=\frac{d\tau}{d\eta}=\frac{1}{a}\gamma\qquad u^i=\frac{dx^i}{d\eta}=\frac{dx^i}{d\tau}\frac{d\tau}{d\eta}=\frac{1}{a}\gamma v^i
\end{align}
where $\gamma=1/\sqrt{1-v^2}$.

The (square root of the) determinant and the inverse of \eq{bmatrixfrw} are then given by 
\begin{equation}
\label{eqforb}
b=\frac{1}{a^3}\det(\partial\Phi)\sqrt{1-v^2}
\end{equation}
\begin{equation}
\label{Binverse}
B^{-1}=a^2\partial\Phi^{-1}\left( \II+\frac{1}{1-v^2} v\otimes v \right)(\partial\Phi^T)^{-1}\,.
\end{equation}
The equilibrium solution of the fluid in the FLRW  background is given by $\Phi^i=x^i$ and therefore
 \begin{equation}
\bar{B}^{ij}=\frac{1}{a^2}\delta^{ij}\qquad  \bar{b}=\frac{1}{a^3}\qquad \bar{v}^i=0\,.
\end{equation}
Moreover, the conserved currents can  be explicitly expressed in a simple form
\begin{align}
\mathcal{J}^0_{(\varepsilon)}=& (\rho+p)\varepsilon^a(\Phi) \frac{\partial x^i}{\partial\Phi^a} \frac{v^i}{1-v^2}\\ 
\mathcal{J}^i_{(\varepsilon)}=& (\rho+p)\varepsilon^a(\Phi)\frac{\partial x^j}{\partial\Phi^a}\left(\delta_j^i+\frac{v^i v^j}{1-v^2}\right)\,,
\end{align}
which is suitable for a Lagrangian formulation (see appendix \ref{appEulerLagrange}) of the conserved charges
\begin{equation}
\label{Qepsilon}
Q_{(\varepsilon)}=-\int d^3x\, a^3\mathcal{J}^\mu_{(\varepsilon)} n_\mu=\int \frac{d^3\Phi}{b}(\rho+p)\varepsilon^a(\Phi) \frac{\partial x^i}{\partial\Phi^a} u_i
\end{equation}
where $n^\mu$ is the unit normal vector to constant time hypersurfaces.
As expected, the background carries, apart from the entropy,  no charges: $\bar{n}_\mu \bar{\mathcal{J}}^\mu_{(\varepsilon)}=0$.
The ground state configuration is thus neutral but supports charge excitations.

Notice that among the various solutions of the transversality constraint (\ref{epsilondiff}) there exists a class \cite{Dubovsky:2005xd}, $\varepsilon^a(\Phi)=\alpha_c \epsilon_{abc}\partial_b  \delta^3(\Phi-\tilde{\Phi})$ with  $\tilde\Phi$ and $\alpha_c$ arbitrary constants, such that the integral in (\ref{Qepsilon}) gives rise to 
a time independent exact 2-form $\Omega$ (living in the internal three-dimensional $\Phi$-space):
\begin{equation}
\label{vorticityforms}
\Omega=dV \qquad V=\left(\frac{\rho+p}{b}\right)u_i \frac{\partial x^i}{\partial \Phi^a}d\Phi^a\qquad
 \dot\Omega=\dot Q=0\,.
\end{equation} 
The conserved charges associated to this class of solutions of (\ref{epsilondiff})  are nothing but the coordinates of the dual 1-form:  $\star\Omega=Q_a d\Phi^a$.
In turn, the circulation $\mathcal{V}$ defined as the \textit{flux} of $Q$ over an arbitrary surface $\mathcal{S}$ in the internal $\Phi$-space 
\begin{equation}
\label{vorticitydef}
 \mathcal{V}  \equiv \int_{\mathcal{S}} \Omega=\oint_{\partial\mathcal{S}} V_a d\Phi^a = \oint_{\Gamma=x(\tau,\partial\mathcal{S})} \left(\frac{\rho+p}{b}\right) u_i dx^i 
\end{equation}
 is conserved on-shell
 \begin{equation}
 \dot{\mathcal{V}}=0\,.
 \end{equation}
 This result represents the generalization to FLRW metrics of the standard non relativistic Kelvin's circulation theorem. Notice that the factor $(\rho+p)/b$ reduces to $m n/s$ in the non relativistic limit for a system of particles of equal mass $m$ and number density $n$, where $s=b$ is the entropy density (as discussed in the previous section). 

\section{Perturbations}
\label{perturb1}

In this section we present the basic set-up for the study of fluid perturbations within this formalism. We will now focus on fluid perturbations alone and include metric fluctuations in Section \ref{metricp}. For simplicity, we assume that there is a single fluid, since the generalization to an arbitrary number of non-interacting components is straightforward. The starting point is the background state\footnote{There actually exist infinitely many background states $\Phi^i=\lambda x^i$ defined by the  constant (in space and time) parameter $\lambda$ which fixes the background value of $\bar{b}=\lambda^3/a^3$. 
Instead of keeping track of all the various factors of $\lambda$ by expanding around one of these vacua, e.g.  by using  $\Phi^i=\lambda(x^i+\pi^i_{(\lambda)})$, we simply work with $\lambda=1$ and recover the general expressions valid also for $\lambda\neq 1$ by expressing the physical results in terms of the corresponding generic background values $\bar{b}$, $\bar{\rho}=-F(\bar{b})$, $\bar{F}_{b}=F_b(\bar{b})$, etc. For example, two fluids described by the same function $F$ but different backgrounds $\bar{b}$'s, depending on the values of the $\lambda$'s, will support perturbations with different speeds of sound which, nevertheless, are still expressed by (\ref{sound}) that is just evaluated into different background values.}, 
 $  \Phi^i(\tau,x)=x^i$, to which we add small independent fluctuations $\pi^a$ in each space direction
\begin{align} \label{lcoord}
\Phi^a=x^a+\pi^a\,.
\end{align} 
The natural expansion parameters are $\dot\pi$ and $\partial \pi$.
Physically, this means that the actual fluid cannot be described only using the background comoving coordinates because there are small inhomogeneities at each point which effectively break the symmetry properties of the system. Recalling the introduction in Section \ref{effective}, this assignment of coordinates corresponds to switching from the isomorphism between the system of coordinates $\mathcal{O}$ and the unperturbed background fluid $\mathcal F$, to another one in which the target space is now the actual imperfect fluid. 

We have
\begin{equation} \label{wth}
\partial\Phi=\II+\partial\pi\qquad \dot\Phi=\dot\pi\qquad \det \partial\Phi=1+\partial_i \pi^i-\frac{1}{2}\partial_i\pi^j \partial_j\pi^i+\frac{1}{2}(\partial_i \pi^i)^2+\ldots
\end{equation}
and expanding the determinant of the matrix $B$ up to second order in $\pi$ we get 
\begin{equation} \label{bseries}
b=\bar{b}\left(1+\partial_i\pi^i-\frac{1}{2}\dot\pi^2-\frac{1}{2}\partial_i\pi^k\partial_k\pi^i+\frac{1}{2}\partial_i\pi^i \partial_j\pi^j+\ldots\right)
\end{equation}
where $\bar{b}=a^{-3}$.
Therefore, after eliminating total derivatives, the action for the perturbations is
\begin{align} \label{act}
S_\pi=\int d^4x\, \sqrt{-g}\,\left(\mathcal{L}_\pi^{(2)}+\ldots\right)
\end{align}
with
\begin{equation}
\label{pert}
\mathcal{L_\pi}^{(2)}=-\frac{\bar b\bar{F_b}}{2}\left(\dot\pi^2 -c_s^2(\partial_i\pi^i)^2\right)
\end{equation}
being the quadratic part of the Lagrangian for perturbations, where we have defined the speed of sound:
\begin{equation} \label{sound}
c_s^2=\bar{b}\frac{\bar{F}_{bb}}{\bar{F}_b}\,.
\end{equation}
In the case of dust $c_s=0$, whereas for radiation the speed of sound reaches, as expected, the value $c_s^2=1/3$. The no-ghost condition  is simply $\bar{b}\bar{F}_b<0$\,. Recalling (\ref{trace}),  this inequality is equivalent to $\bar{\rho}+\bar{p}>0$. Imposing also that $c_s^2\geq0$ gives in turn  $\bar{F}_{bb}\leq0$. 
Notice that at this level, neglecting metric fluctuations, the dynamics of the perturbations (as well as that of the background) is entirely given in terms of $\bar{F}$, $\bar{F}_b$ and $\bar{F}_{bb}$. Higher order derivatives enter only in the self-interaction terms. 

Let us also point out that the contribution of $\mathcal L_\pi^{(1)}$ to the action, the linear piece of the Lagrangian $\mathcal L_\pi$, vanishes because it gives rise to a total derivative that we can omit in \eq{act}, since the background solution depends just on time by construction.

It is straightforward to link the four-velocity of the fluid to the $\pi$ fields that we have introduced in \eq{lcoord}. This can be done at any desired order in $\pi$ by solving iteratively the equations \eq{fluidvelocity}, which are sufficient to determine the four-velocity up to a sign. At second order in $\pi$ (and choosing $u^0$ to be positive) the result is 
\begin{align} \label{5vel}
u^0=\frac{1}{a}\left(1+\frac{1}{2}\dot\pi^2\right) \qquad u^i=\frac{1}{a}\left(-\dot\pi^i+\dot\pi^k\partial_k\pi^i\right)\,.
\end{align}
Notice that $v$, as defined in \eq{velo}, can be expanded as 
\begin{align}
v=-(\II-\partial\pi)\dot\pi+\ldots
\end{align}
which is consistent with the expression $v^i=a\, u^i$\,, by comparison with \eq{5vel}. From the last expression we recover \eq{5vel} by neglecting $\gamma$, which is irrelevant up to third order in $\pi$.

In the following, we shall discuss the transverse  and longitudinal modes of the Lagrangian (\ref{pert}) by decomposing  the coordinate perturbation:
\begin{equation} \label{comps}
\pi=\pi_L +\pi_\perp\qquad \nabla\cdot \pi_\perp=0\qquad \nabla\times\pi_L=0\,
\end{equation}
such that the second order Lagrangian for $\pi$ can be written as
\begin{align} \label{hpert}
\mathcal{L_\pi}^{(2)}=\frac{1}{2}(\bar{\rho}+\bar{p})\left(\dot\pi_\perp^2+\dot\pi_L^2-c_s^2(\nabla\cdot\pi_L)^2\right)\,.
\end{align}

The generalization to $N$ fluids with different density functions $F_\alpha$ $(\alpha=1,\ldots,N)$ is simple. Each fluid will be described by internal coordinates $\Phi_\alpha^a$ which can be perturbed independently using  different $\pi_\alpha^i$\,. If we assume that the fluids do not interact directly among themselves, which means that the Lagrangian does not contain terms mixing different sets of $\Phi_\alpha$ coordinates, the generalization of \eq{hpert} is just
\begin{align} \label{hpertN}
\mathcal{L_\pi}^{(2)}=\frac{1}{2}\,\bar\rho_T\sum_{\alpha=1}^N (1+w_\alpha)\Omega_\alpha \left(\dot\pi_{\alpha\perp}^2+\dot\pi_{\alpha L}^2-c_{\alpha s}^2(\nabla\cdot\pi_{\alpha L})^2\right)
\end{align}
where we define the relative background energy density of each fluid as the following time function:  $\Omega_\alpha=\bar{\rho}_\alpha/\bar{\rho}_T$\,. As we will see later, when we introduce metric perturbations, the specific case in which the $\pi_{\alpha L}$ are the same for two or more fluids is of particular interest because it describes adiabatic perturbations between those species. 
We end this section writing the equations of motion for transverse\footnote{The equation~(\ref{eomtransverse}) corresponds to the equation of vorticity conservation, see equation (\ref{vorticityforms}). In fact, at this order in $\pi$, we have $\star\Omega=-\nabla \times (a^4(\bar{\rho}+\bar{p})\dot{\pi}_{\perp})$.}
\begin{equation} \label{eomtransverse}
\frac{d}{d\tau}\left(a^4(\bar\rho_\alpha+\bar p_\alpha) \dot\pi_{\alpha \perp}^i \right)=0 
\end{equation}
and longitudinal modes
\begin{align} \label{eomlong}
\ddot\pi_{\alpha L}^i+(1-3\,c_{\alpha A}^2)\ch\dot\pi_{\alpha L}^i -c_{\alpha s}^2\partial_i(\nabla\cdot\pi_{\alpha L})=0  \qquad {i=\{1,2,3\}}\qquad \alpha=1,\ldots,N\,.
\end{align}
In this equation $c_{\alpha A}^2$ represents the standard {\it adiabatic} speed of sound of the fluid with label $\alpha$, which for this type of fluids is simply the ratio between the time variation of the background pressure and energy density:
\begin{align} \label{adiab}
\dot{\bar p}_\alpha = c_{\alpha A}^2\,\dot{\bar\rho}_\alpha\,.
\end{align}
In general, the adiabatic speed of sound is defined in hydrodynamics as the derivative of the pressure with respect to the density. In our case:
\begin{align}
c_{\alpha A}^2\equiv\left. \frac{d p_\alpha}{d \rho_\alpha}\right|_{b=\bar{b}}\,.
\end{align}
This definition reduces to \eq{adiab} for a fluid given by \eq{lagrangian}. Using the equations (\ref{rho}) and (\ref{trace}), it is straightforward to check that, in this case, the adiabatic speed of sound $c_{\alpha A}^2$ actually coincides with the speed of sound $c_{\alpha s}^2$ that we introduced in (\ref{sound}):
\begin{align} \label{equalsound}
c_{\alpha A}^2=c_{\alpha s}^2
\end{align} 
for any fluid of the form \eq{lagrangian}. This shows that the only possible speed of sound of a perfect fluid defined by the Lagrangian (\ref{lagrangian})  is adiabatic. As we will see later,  the longitudinal modes, which (in absence of metric perturbations) are given by the evolution equation \eq{eomlong}, are intrinsically related to the adiabatic modes of the fluid.

If we use the expression (\ref{bseries}) to expand at first order the energy density and the pressure of a fluid defined by \eq{lagrangian},  we find that the corresponding perturbations are
\begin{align} \label{matteronly}
\delta_M \rho =(\bar\rho+\bar p)\, \nabla \cdot\pi_L \qquad \delta_M p= (\bar\rho+\bar p)\,\frac{\bar{b}\,\bar{F}_{bb}}{\bar F_b}\, \nabla\cdot\pi_L
\end{align}
in agreement with \eq{equalsound}. In these formulas we have introduced the subscript $_M$ to indicate that these are the perturbations coming exclusively from the variation of the matter part of the action \eq{action}. In the next section we include metric fluctuations to achieve a complete description of the perturbations. The equations 	\eq{matteronly} will turn out to be modified by a correction coming from the metric inhomogeneities; see equation \eq{mattmet}.

\section{Including metric perturbations} \label{metricp}

Assuming that the theory of gravity is General Relativity\footnote{The formalism can also be applied to theories of modified gravity.} (GR) the full action takes the form
\begin{align} \label{fulla}
S=S_{EH}+S_m
\end{align}
where the first piece is the usual Einstein-Hilbert action
\begin{align}\label{EH}
S_{EH}=\frac{1}{16\pi G}\int d^4 x\sqrt{-g} R
\end{align}
and the second one is the matter part of $S$, which for each perfect fluid of the type we have been studying is given by \eq{action}.

Let us now focus on $S_m$\,, with the aim of  finding the terms that constitute the direct interaction between metric and matter perturbations. A straightforward way of doing this at any desired order is to expand the action $S_m$ using a functional series. We expand first with respect to the metric variations taking an arbitrary matter configuration
\begin{align}\label{expansion}
S_m[\varphi, g_{\mu\nu}+\delta g_{\mu\nu}] =& S_m[\varphi, g_{\mu\nu}]+\frac{1}{2}\int d^4x \sqrt{-g}\,T^{\mu\nu}\delta g_{\mu\nu}(x)+\ldots
\end{align}
where $\varphi$ globally represents all matter fields (e.g. dark matter, dark energy, etc) and $T^{\mu\nu}$ is the gravitational energy-momentum tensor.
Expanding now the equation \eq{expansion} around a matter background we end up with the following action for the matter-gravity coupling at linear order in metric perturbations (but all orders in the matter fields)
\begin{equation} \label{contains}
S_m\supset\frac{1}{2}\int d^4x \sqrt{-g}\,\delta_M T^{\mu\nu}\delta g_{\mu\nu}(x)
\end{equation}
where $\delta_M T^{\mu\nu}$ is the variation of $T^{\mu\nu}$ induced by the matter perturbations. The computation of the matter-metric mixing terms is then very simple provided that we know  the form of the energy-momentum tensor. In particular,  for a perfect fluid \eq{lagrangian}, this is just given by \eq{pf}. We will now see in detail how this works in the conformal Newtonian gauge.

In the following part of this section we focus only on scalar perturbations that mix with longitudinal modes.
Vector and tensor metric perturbations will be discussed in Section \ref{vectorperturb}. 
In the conformal Newtonian gauge the perturbed FLRW metric is then diagonal:
\begin{align} \label{conformal}
ds^2=a^2\left(-(1+2\psi)d\tau^2+(1-2\phi)\delta_{ij}dx^idx^j\right)
\end{align}
If the universe did not contain any imperfect fluids at all, we would have $\psi=\phi$ at linear order in the equations of motion. However, we want to include the more general possibility that some fluids with anisotropic stress could also be present and therefore we will treat these two potentials as distinct variables. This, for instance,  happens at very early times when the anisotropic stress of neutrinos cannot be neglected. 

Let us consider the energy momentum tensor of some species in their rest frame. Using that $u^2=-1$\,, we find the following expression for the (lowest order in metric fluctuations) coupling between matter and metric perturbations (of scalar type)
\begin{align} \label{mixednew}
\frac{1}{2}\sqrt{-g}\,\delta_M T^{\mu\nu}\delta g_{\mu\nu}=-a^4(\psi\,\delta_M\rho+3\phi\,\delta_M p)
\end{align}
where the fluctuations in the energy density and the pressure include all orders. Notice that for a perfect fluid defined by \eq{lagrangian}  this coupling gives the following contribution to the action for fluctuations at first order in the metric perturbations
\begin{align} \label{mixact}
S_m\supset-\int d^4x\, a^4(\psi+3c_s^2\phi)\bar{\rho}\,\delta_M
\end{align}
where $\delta_M\equiv \delta_M\rho/\bar{\rho}$  denotes the intrinsic (not metric) relative energy density perturbation at all orders and $c_s^2$ is given by \eq{sound}. For linear matter perturbations of such a fluid we can use \eq{matteronly} and therefore the mixing between matter and metric perturbations is
\begin{equation} \label{mixt}
S_m\supset S_{m-g}^{(2)}=-\int d^4x\, a^4(\bar\rho+\bar p)(\psi+3c_s^2\phi)\,\partial_i\pi^i_L
\end{equation}
which shows that metric fluctuations and transverse modes do not couple at this order.

As a  check of these results, we perform an explicit  independent calculation by writing the square root $b$ and $\sqrt{-g}$ in the perturbed metric \eq{conformal}. The matrix $B$ has coefficients
\begin{align}
a^2B^{ij}=-(1+2\psi)^{-1}\dot\Phi^i\dot\Phi^j+(1-2\phi)^{-1}\partial_k\Phi^i\partial_k\Phi^j
\end{align}
and the square root of its determinant is
\begin{equation}
\label{b_metricperturbed}
b=\frac{1}{a^3}\frac{1}{(1-2\phi)^{3/2}}\det(\partial\Phi)\sqrt{1-\frac{1-2\phi}{1+2\psi}v^2}
\end{equation}
where we recall that the definition of the coordinate velocity is \eq{velo} and $\det(\partial\Phi)$ is still given by \eq{wth}.
In addition, the metric determinant is 
\begin{align}\label{deternewt}\sqrt{-g}=a^4\sqrt{(1+2\psi)(1-2\phi)^3}\end{align}
so that the mixing term at linear order in the metric perturbations is
\begin{equation}
\sqrt{-g}\mathcal{L}_m\supset a^4\,(\psi-3\phi)\,\delta F+3a^4\,\phi\,\delta(b F_b)
\end{equation}
which, using \eq{rho} and \eq{trace}, is in agreement with the formulas above. 

We can easily compute all (metric and matter) second order terms that come from the matter action \eq{action} by simply doing a Taylor series expansion. The result is
\begin{align}\nonumber \label{mstaylor}
 S_m^{(2)}= \,&\frac{1}{2}\int  d^4x\, a^4\,(1+w)\,\bar\rho\,\left[ \dot\pi^2-c_s^2(\partial_i\pi^i_L)^2 - 2(\psi+3c_s^2\phi)\,\partial_i\pi^i_L\right]\\
 &+\frac{1}{2}\int d^4x\,a^4\,\bar\rho \left[\psi^2+3(w-3(1+w)c_s^2)\phi^2-6w\phi\,\psi\right]\,.
 \end{align}
The first line in this expression corresponds to the mixing term \eq{mixt} that we have just computed plus the piece that comes from \eq{hpert}, which is the purely matter part of the action that we already obtained by neglecting metric perturbations. The second line of \eq{mstaylor} is the contribution of the matter action \eq{action} to the action of the metric fluctuations. In order to get the full action for perturbations at second order we have to complete the metric part by perturbing the Einstein-Hilbert action \eq{EH}. This can be found, for any gauge, in \cite{Mukhanov:1990me}. 

The generalization of  \eq{mstaylor} to  $N$ fluids of this kind, interacting only via gravity, is straightforward. We just have to sum the individual actions of the different components. Then, we can easily write down the equations of motion for the transverse and longitudinal modes of each component. The transverse modes are unaffected by the scalar metric perturbations at this order and their equations of motion are still given by \eq{eomtransverse}:
\begin{equation} \label{eomtransverse2}
\frac{d}{d\tau}\left(a^4(\bar\rho_\alpha+\bar p_\alpha) \dot\pi_{\alpha\perp} \right)=0 \qquad \alpha=1, \ldots, N\,.
\end{equation}
On the other hand, the equation \eq{eomlong} for the longitudinal modes gains a contribution from the metric perturbation, becoming:
\begin{align} \label{elongnew}
 \ddot{\pi}_{\alpha L}+\mathcal{H}(1-3c_{s_\alpha}^2)\dot\pi_{\alpha L}-c_{s_\alpha}^2\nabla(\nabla\cdot \pi_{\alpha L})-\nabla\psi-3c_{s_\alpha}^2\nabla\phi=0 \qquad \alpha=1, \ldots, N\,.
  \end{align}

The expressions in \eq{matteronly} for the density and pressure perturbations get modified when metric fluctuations are included. Concretely, they are replaced by\footnote{This expression for the energy density reminds of the Zel'dovich approximation,  $\delta\rho\sim\nabla\cdot\pi$, in Newtonian Lagrangian perturbation theory (see for example \cite{Bernardeau:2001qr}), with $\pi$ playing the role of the displacement field. The extra metric perturbation term $\phi$ in \eq{mattmet} comes from our relativistic formulation.}
\begin{align} \label{mattmet}
\delta \rho_\alpha =(\bar{\rho}_\alpha+\bar{p}_\alpha)(\nabla\cdot\pi_\alpha+3\phi) \qquad \delta p_\alpha=c_{\alpha s}^2\,\delta\rho_\alpha
\end{align}
with the speed of sound squared $c_{\alpha s}^2$ defined exactly as before, in \eq{sound}. Notice that both the density and pressure perturbations gain a term that depends on the metric potential $\phi$. In particular, for the total (matter plus metric) density perturbation of each fluid we write
\begin{align} \label{decomp}
\delta\rho_\alpha=\delta_M\rho_\alpha+3(\bar\rho_\alpha+\bar p_\alpha)\phi
\end{align}
where the matter part $\delta_M\rho_\alpha$ is given by the first expression of \eq{matteronly}.

The four-velocities of the fluids also change with respect to \eq{5vel} due to the effect of the metric perturbations. The equations \eq{fluidvelocity} are both still valid and we can again find the four-velocity solving them iteratively. Alternatively, proceeding in the same way that we have used to obtain \eq{gammavel} we get
\begin{align}
u^0=\frac{1}{a}\,\tilde\gamma\qquad u^i=\frac{1}{a}\,\tilde\gamma\, v^i \qquad
\tilde\gamma^{-2}\equiv(1+2\psi)-(1-2\phi) v^2
\end{align}
where the coordinate velocity is still given by
\begin{align} \label{vgen}
v^i=\frac{d x^i}{d\tau}=-\dot\pi^i+\dot\pi^k\partial_k\pi^i+\ldots
\end{align}
which is the same that we would derive from \eq{5vel} (where metric perturbations are set to zero).  This happens because $v^i$ is the solution of \eq{vsolver}, which is the same with or without metric perturbations. Expanding $\tilde\gamma$ at second order in perturbations we get
\begin{align} \label{cuadvelex}
u^0=\frac{1}{a}\left(1-\psi+\frac{3}{2}\psi^2+\frac{1}{2}v^2+\ldots\right) \qquad u^i=\frac{1}{a}\left(1-\psi+\ldots\right)v^i\,.
\end{align}

Let us point out that the expression \eq{cuadvelex} for $u^i$ differs (at second order) from the one found in the review \cite{Malik:2008im} and other related papers in the literature. 

In the next section, we will use the equation \eq{elongnew} to understand adiabatic modes. In order to do so, we will need to replace $\phi$ and $\psi$ by their expressions in terms of pure  matter perturbations. The difference between the two metric potentials is given, at first order, by the total anisotropic stress of the fluid system. If all its components are perfect, the metric potentials are equal at first order. Any difference between the metric potentials will then be due to imperfect components. Another useful piece of information is the (general relativistic) Poisson equation, which allows to express the Laplacian squared of $\phi$ as a function of the density and velocity perturbations. This equation can be derived by combining the $0-0$ and $0-i$ Einstein equations (see for instance \cite{Baumann:2010rc}) or from the variation of the second order perturbed action \eq{fulla} with respect to a metric degree of freedom that is later set to zero in the conformal Newtonian gauge \cite{Mukhanov:1990me}.  The Poisson equation reads:
\begin{align} \label{psnf}
\nabla^4\phi=\frac{3}{2}\ch^2\sum_{\alpha=1}^N\Omega_\alpha\left(\nabla^2\delta_\alpha-3\left(1+w_\alpha\right)\ch \theta_\alpha\right)
\end{align}
where we sum over all fluid species and $\theta_\alpha$ is the standard notation for the divergence of the velocity perturbation $v_\alpha$ \cite{Ma:1995ey}\,, as given in \eq{vgen}:
\begin{align} \label{theta}
\theta_\alpha=\nabla\cdot v_\alpha=-\nabla\cdot \dot{\pi}_\alpha+\ldots
\end{align}
Since for a perfect fluid, $\delta_\alpha$ is given by \eq{mattmet}, it turns out that the right-hand side of \eq{psnf} contains a contribution that depends on the spatial metric perturbation $\phi$. Hence the expression \eq{psnf} becomes a partial differential equation for $\phi$ with second and fourth order derivatives. We will use such a form of the Poisson equation in Section \ref{genad} when discussing adiabatic perturbations, see equation~(\ref{adiab3}).

Before moving to the description of adiabatic modes, it is interesting and useful to see how the equations of motion and \eq{eomtransverse2} and \eq{elongnew}  are related to the standard continuity and Euler equations for generic fluids that have the following general form in the conformal Newtonian gauge \cite{Ma:1995ey}:
\begin{align} \label{continuity}
\dot\delta_\alpha &= -(1+w_\alpha)(\theta_\alpha-3\dot\phi)-3\left(\frac{\delta P_\alpha}{\delta \rho_\alpha}-w_\alpha\right)\ch\delta_\alpha\\ \label{euler}
\dot\theta_\alpha &=-\left(1-3w_\alpha\right)\ch\theta_\alpha -\frac{\dot{w_\alpha}}{1+w_\alpha}\theta_\alpha-\frac{1}{1+w_\alpha}\frac{\delta P_\alpha}{\delta \rho_\alpha}\,\nabla^2 \delta_\alpha-\nabla^2\psi+\nabla^2\sigma_\alpha\,.
\end{align}
Notice that for these fluids, i.e. described by (\ref{lagrangian}), the Euler equation (\ref{euler}) with $\sigma_\alpha=0$ is precisely the divergence of (\ref{elongnew}), while the continuity equation (\ref{continuity}) is just an identity.
This follows from the fact that the velocity perturbations $\theta_\alpha$ in (\ref{theta}) are the time derivative of the longitudinal modes $\nabla\cdot\pi_\alpha$ rather than independent variables.

\subsection{Vector and tensor metric perturbations}
\label{vectorperturb}

We have derived the equations of motion \eq{eomtransverse2} and \eq{elongnew} for an ensemble of $N$ non-interacting fluids, each with its own $\pi_\alpha$ field, assuming only scalar metric perturbations. However, since $\pi_{\alpha\perp}$ can be written for each fluid as the curl of a vector potential, we can expect its dynamics to be affected by metric vector perturbations. For the longitudinal component $\pi_{\alpha L}$ this cannot occur because it has zero curl and hence it can be expressed as the gradient of a single scalar degree of freedom, which then couples only to $\phi$ and $\psi$ as we have already seen. 

The perturbed FLRW metric in the Poisson gauge \cite{Bertschinger:1993xt,Bruni:1996im} generalizes the conformal Newtonian gauge \eq{conformal} to include vector and tensor degrees of freedom:
\begin{align} \label{poissongauge}
ds^2=a^2\left(-(1+2\psi)d\tau^2+2\nu^i \,d\tau dx^i+\left[(1-2\phi)\delta_{ij}+\chi_{ij}\right]dx^idx^j\right)
\end{align}
where $\nu$ and $\chi$ are respectively pure vector and tensor perturbations that satisfy
\begin{align}
\partial_k \nu^k=0\qquad\partial_k\chi_{kj}=0\qquad \chi_{jj}=0\,.
\end{align}
Applying \eq{contains} to the full set of metric perturbations (transverse) $\nu$ vector  and (transverse and traceless) $\chi$ tensor  metric degrees of freedom we find
\begin{align}
\label{allperturbations}
\frac{1}{2}\sqrt{-g}\delta_M T^{\mu\nu}\delta g_{\mu\nu}= & a^4\left[-(\psi\,\delta_M\rho+3\phi\,\delta_M p)+a(\bar\rho+\bar p)\nu^i \delta_M u^i \right]\\
=&-a^4(\bar\rho+\bar p)\left[(\psi+3c_s^2\phi)\,\partial_i \pi^i_L +\dot{\pi}^i_T \nu^i\right]+\mbox{(total derivative)}
\end{align}
where tensor perturbations do not appear because of the tracelessness condition. This confirms the known result that tensor metric perturbations do not couple, at the lowest order in perturbation theory, to perfect fluids.
On the other hand, the transverse modes (vortices) mix with the vector metric perturbations. The equations \eq{eomtransverse2}  get now replaced by a conservation equation of the linear combination $(\nu^i-\dot{\pi}_T)$
\begin{align} \label{vecttrans}
\frac{d}{d\tau}\left(a^4(\bar\rho_\alpha+\bar p_\alpha) \left(\nu^i-\dot\pi^i_{\alpha\perp}\right) \right)=0 \qquad \alpha=1, \ldots, N
\end{align}
where, as before, the index $\alpha$ lists the different perfect fluids from 1 to $N$. 
This equation is nothing but the conservation of the charges associated with the vorticity, see (\ref{Qperturb}).
It corresponds also to the separate conservation of the three-momentum for each fluid \cite{Malik:2008im,Bardeen:1980kt}:
\begin{align}  \label{mcon}
\delta \dot q^j_\alpha+4\ch\delta q^j_\alpha=0\qquad \alpha=1, \ldots, N
\end{align}
where 
\begin{align}
\delta q^i_\alpha\equiv(\bar\rho_\alpha+\bar p_\alpha)(\nu^i-\dot\pi^i_{\alpha\perp}) \qquad \alpha=1, \ldots, N
\end{align}
are the three-momenta of the fluids. Notice that, if the fluids were not perfect, the right-hand side of \eq{mcon} (or equivalently \eq{vecttrans}) would have an anisotropic stress source term. The equation \eq{mcon} tells us that in absence of such a term, the three-momentum decays due to the Hubble expansion.

The equation for the evolution of the vector metric perturbations is \cite{Bardeen:1980kt}
\begin{align}
\nabla^2\nu^i=6\ch^2\sum_\alpha(1+w_\alpha)(\nu^i-\dot\pi^i_{\alpha\perp})
\end{align}
consistent with the redshift of the three-momenta at large scales. 

For completeness we include the equation for tensor modes, which in absence of any source of anisotropic stress reads:
\begin{align}
\nabla^2\chi_{ij}=\ddot\chi_{ij}+2\ch\dot\chi_{ij}\,.
\end{align}
As it is well known, these modes do not couple to any fluid degree of freedom if there is no anisotropic stress.

\section{Adiabatic modes} \label{genad}

Adiabatic perturbations correspond to modes associated with equal time shifts. They are therefore constructed by perturbing any homogeneous (intensive) fluid variable $\mathcal I_j$ in the following way
\begin{align} \label{adiabm}
\bar{\mathcal{I}}_j(\tau)\rightarrow \mathcal{I}_j(\tau,x)=\bar{\mathcal{I}}_j(\tau+\hat{\pi}(\tau,x))=\bar{\mathcal{I}}_j(\tau)+\hat{\pi}(\tau,x)\dot{\bar{\mathcal{I}}}_j+\ldots
\end{align}
using the same $\hat\pi$ for all $\mathcal I_j$\,, where the subscript $_j$ lists any intensive variable pertaining to the fluid. Therefore, we say that two intensive fluid variables $\mathcal{I}_1$ and $\mathcal{I}_2$ are adiabatically related if their perturbations can be constructed from the same time shift $\hat\pi$. If we focus on the energy density $\rho$ and the pressure $p$ of any fluid, at first order in $\hat\pi$ this implies
\begin{align} \label{vels}
c_{ A}^2=c_{ s}^2\,.
\end{align}
Moreover, for several fluids, we have that adiabatic density modes satisfy
\begin{align}
\frac{\delta\rho_\alpha}{\dot{\bar{\rho}}_\alpha}=\frac{\delta\rho_\beta}{\dot{\bar{\rho}}_\beta}
\end{align}
for any pair of fluids $\alpha$ and $\beta$ interacting only through gravity. Describing the perturbations as in \eq{lcoord},
we get that the adiabatic mode is nothing but a common longitudinal degree of freedom
\begin{align} \label{timeshi}
\hat\pi=\frac{\delta\rho_\alpha}{\dot{\bar{\rho}}_\alpha}=\frac{\delta p_\alpha}{\dot{\bar{p}}_\alpha}=-\frac{1}{3\mathcal{H}}\left(\nabla\cdot\pi+3\phi\right)\quad \forall \alpha\,.
\end{align}
In other words, the condition of adiabatic modes translates into ``flavour'' (or species) independence of the longitudinal modes
\begin{equation} \label{adbc}
\pi_{\alpha L}=\pi_{\beta L}\,.
\end{equation}
In the next subsection we show how this result extends to imperfect fluids that have vanishing non-adiabatic pressure perturbations.

For convenience, we use the curvature perturbation on uniform density hypersurfaces, which is defined to be \cite{Bardeen:1983qw}
\begin{align} \label{zbard}
\zeta_\alpha=-\phi-\ch\frac{\delta\rho_\alpha}{\dot{\bar{\rho}}_\alpha}\,.
\end{align}
Remarkably, in our case this is just one third of the divergence of $\pi_\alpha$
\begin{align} \label{zeta}
\zeta_\alpha=\frac{1}{3}\nabla\cdot\pi_\alpha
\end{align}
and taking its derivative with respect to conformal time, we get the following relation with the divergence of the velocity (at first order in $\pi$):
\begin{align} \label{rvz}
\theta_\alpha+3\dot\zeta_\alpha=0\,.
\end{align}
Defining the entropy perturbation between two species in the usual way \cite{Kodama:1985bj}
\begin{align}
S_{\alpha\beta}=3(\zeta_\alpha-\zeta_\beta)
\end{align}
we see that for the kind of fluids that we consider, the entropy perturbation among two species is zero if the coordinate perturbations $\pi_\alpha$ $(\alpha=1,2$) of the two fluids differ by at most a divergenceless three-vector. This is exactly the condition \eq{adbc} for adiabatic modes that we have found before.

\subsection{Adiabatic modes for imperfect fluids}
Let us recall that for more general (imperfect) fluids\footnote{For a study of cosmological perturbations of imperfect fluids, in particular in connection to scalar fields, see \cite{Sawicki:2012re}.} entropy modes may come as well from an intrinsic non-adiabatic pressure perturbation. The total pressure perturbation of any species can be decomposed as
\begin{align} \label{nadpress}
\delta P_\alpha=\delta P_{\alpha (nad)}+c_{\alpha A}^2\delta\rho_\alpha
\end{align}
where the second term is the product of the density perturbation and the usual adiabatic speed of sound.
For perfect fluids like the ones we have studied in this work, $\delta P_{\alpha(nad)}$ (the non-adiabatic part of the pressure perturbation) is zero. However, when a fluid has other internal degrees of freedom different from the $\Phi$ coordinates\footnote{See Appendix \ref{appa} for the general form of the energy-momentum tensor in this case.}, an intrinsic non-adiabatic pressure will typically arise. 
Nevertheless, there exists an interesting class of imperfect fluids that have vanishing non-adiabatic pressure perturbations. They fail to be perfect only because they have anisotropic stress $\sigma$
\begin{align}
\sum_\alpha(\bar\rho_\alpha+\bar p_\alpha)\nabla^2\sigma_\alpha=\sum_{i, j}\partial_i\partial_j T^i_j-\frac{1}{3}\sum_k\nabla^2T^k_k\,,
\end{align}
that enters in the Einstein equations for the scalar potentials
\begin{align} \label{adiab4}
\nabla^4(\psi-\phi)=\frac{9}{2}\mathcal{H}^2\sum_{\alpha}\Omega_\alpha (1+w_\alpha)\nabla^2\sigma_\alpha\,.
\end{align}
At early times, neutrinos fall into this class of fluids: $\sigma_\nu\neq0$ while $\delta P_{\nu (nad)}=0$.

As we are going to see next, in an adiabatic mode, the expressions that occur for the density and velocity perturbations of perfect fluids are also valid at first order in $\pi$ for imperfect fluids of that kind. From the equation (\ref{cons}) we know that $\delta_\alpha/(1+w_\alpha)$ is a species independent ratio for all fluids involved in an adiabatic mode. Since the definition (\ref{zbard}) can be written as $\delta_\alpha/(1+w_\alpha)=3(\zeta_\alpha+\phi)$ we have that 
\begin{align} \label{adiabmo}
\zeta_\alpha =\zeta_\beta
\end{align}
for any two fluids in an adiabatic mode. Moreover, if at least one perfect fluid is present, we have, thanks to equation \eq{mattmet}, $3\zeta=\nabla\cdot \pi$.
For imperfect fluids with vanishing non-adiabatic pressure:
\begin{align}
\label{thetaflavor}
\frac{d}{d\tau}\left(\frac{\delta_\alpha}{1+w_\alpha}\right)=\frac{\dot\delta_\alpha}{1+w_\alpha}+3\ch(c_{\alpha A}^2-w_\alpha)\frac{\delta_\alpha}{1+w_\alpha}=3\dot\phi-\theta_\alpha
\end{align}
where we have used the continuity equation~\eq{continuity} in the last equality. We thus find that in this case the velocity perturbations are also flavor independent
\begin{equation}
\delta P_{\alpha (nad)}=0 \Longrightarrow \theta_\alpha=\theta_\beta=-3\dot\zeta
\end{equation}
for any two fluids sharing an adiabatic mode. Again, if at least one perfect fluid is present, then $\theta$ is proportional to the time derivative of the longitudinal mode: $\theta=-\nabla\cdot \dot\pi$.
Effectively, in an adiabatic mode, the energy density and the velocity of an imperfect fluid with adiabatic speed of sound are like those of a perfect fluid\footnote{See \cite{Ballesteros:2010ks} for how to generalize the standard adiabatic conditions for fluids with non-adiabatic sound speed.}.

\subsection{Superhorizon perturbations}

We  now describe adiabatic modes in further depth using the variable $\zeta$ introduced in \eq{zeta}. Since adiabatic modes are particularly important for the study of initial conditions in the early universe and at those times the anisotropic stress of neutrinos cannot be neglected,  we are led to consider the generic situation in which the metric potentials $\phi$ and $\psi$ can be different from each other, even at first order in the equations.
Armed with the knowledge of the previous sections, we can write the Poisson equation \eq{psnf} in the following way:
\begin{align}
\label{adiab3}
(\nabla^2-M^2)\nabla^2\phi &=M^2\left(\nabla^2\zeta+3\ch\dot{\zeta}\right)
\end{align}
where we introduce the ``effective mass'' (squared)
\begin{align}
M^2 =\frac{9}{2}\mathcal{H}^2\sum_{\alpha}\Omega_\alpha (1+w_\alpha) \qquad
\end{align}
in which the sum extends over all (perfect and imperfect) fluids that are present in the system.

According to the Euler equation \eq{euler}, for fluids with anisotropic stress but zero non-adiabatic pressure, the divergence of the equation \eq{elongnew} must be replaced by
\begin{align} \label{elongnewanis}
\ddot{\zeta_\alpha}+(1-3c_{s_\alpha}^2)\ch\dot\zeta_\alpha-c_{s_\alpha}^2\nabla^2\zeta_\alpha-\frac{1}{3}\nabla^2\psi-c_{s_\alpha}^2\nabla^2\phi+\frac{1}{3}\nabla^2\sigma_\alpha=0
\end{align}
which, together with equations (\ref{adiab3}) and (\ref{adiab4}), form a close system for the perturbations.
If photons (which have $c_s^2=1/3$ and $\sigma_\gamma=0$) are present, recalling that for adiabatic modes the equation \eq{adiabmo} holds, we obtain that the equation (\ref{elongnewanis}) implies
\begin{align}
\label{firsteq}
\ddot\zeta-\frac{1}{3}\nabla^2\left(\zeta+\psi+\phi\right)= & 0\\
\label{secondeq}
(1-3c_{s_\alpha}^2)\left[\ch\dot\zeta+\frac{1}{3}\nabla^2\left(\zeta+\phi\right)\right]+\frac{1}{3}\nabla^2\sigma_\alpha= & 0\,.
\end{align}
These two equations are actually valid only approximately. The same happens for the condition \eq{adiabmo},  which strictly holds just in the limit of large scales, $\nabla/\ch\rightarrow 0$.
At early times, when the universe is radiation dominated and the anisotropic stress of neutrinos cannot be neglected, the equation \eq{secondeq} implies that $\nabla^2\sigma_\nu=0$\,.  Using the equation (\ref{adiab4}), one would then na\"ively conclude that for an adiabatic mode $\nabla^4(\psi-\phi)=0$\,. These two last conclusions have to be understood as approximately valid expressions in the limit $\nabla/\ch\rightarrow 0$. Otherwise they would imply trivial solutions for $\sigma_\nu$ and the difference of the metric potentials \cite{Ma:1995ey}. In the limit of very large wavelengths, the Poisson equation \eq{adiab3}  simplifies to
\begin{align} \label{poissapp}
-\nabla^2\phi =& 3\ch\dot\zeta +\nabla^2\zeta\,.
\end{align}
If we plug \eq{poissapp} into \eq{secondeq} (which assumes adiabaticity), we obtain $\nabla^2\sigma_\alpha=0$, without having made any assumption about $c_{s_\alpha}^2$. This means that the approximation $\nabla^2\sigma_\nu=0$ that we have found for adiabatic modes is valid at very large scales (and early times, with the universe being radiation dominated). Therefore, as we have anticipated, we conclude that adiabatic modes can only be defined at very large scales and, strictly speaking,  there are no exact adiabatic modes. 

Notice that if the perturbations are regular for very large wavelengths, the Poisson equation helps us to write (\ref{elongnewanis}) in the following way
\begin{equation}
\label{zetafinal}
\ddot{\zeta}+\frac{1}{3}\nabla^2\zeta+2\ch\dot\zeta=\frac{3}{2}\ch^2\sum_\alpha\Omega_\alpha(1+w_\alpha) \sigma_\alpha
\end{equation}
which is a simple equation for $\zeta$ where no metric perturbations appear explicitly. If the condition of regularity at large distances wouldn't apply, the whole equation \eq{zetafinal} takes an overall extra $\nabla^2$ operator on both sides.

To conclude this section, let us recall and show that, as it is expected, the curvature perturbation on uniform density hypersurfaces $\zeta$ coincides with (minus) the comoving curvature perturbation $\cal R$ in the limit of very large scales. The definition of $\cal R$ is (see e.g. \cite{Weinberg:1102255}):
\begin{align} \label{comovcp}
{\cal R}\equiv \phi-\ch \tilde\upsilon
\end{align}
where $\tilde\upsilon$ is a scalar potential that gives the longitudinal part of the fluid coordinate velocity, i.e. $\dot\pi_L=-\nabla \tilde\upsilon$\,. Taking the Laplacian of \eq{comovcp} and using \eq{zeta} we get
\begin{align} \label{rela}
\nabla^2{\cal R}=\nabla^2\phi+3\ch\dot\zeta
\end{align}
For perturbations of very large wavelength, we can use \eq{poissapp} to eliminate the time derivative of $\zeta$ in \eq{rela}, obtaining 
\begin{align}
{\cal R}+\zeta=0
\end{align}
assuming again regularity of the fluctuations at very large distances.

\section{Conclusions and Outlook}
\label{conclusions}
 
We have shown how to obtain the fundamental equations for the cosmological evolution of perfect fluids from symmetries and action principles.  
Although we have chosen to focus our analysis on the case of a FLRW metric (and its perturbations) in General Relativity and non-interacting fluids, the method is of ample generality and can be easily applied to other geometries, theories of modified gravity and can be extended to describe systems which do not interact only via gravity. The formalism presented allows a straightforward description of the dynamics of  the expansion both at the background and perturbation level, allowing us e.g. to easily recover standard results of fluid perturbation theory  \cite{Mukhanov:1990me,Ma:1995ey,Bardeen:1980kt,Kodama:1985bj}.

The method is based on an effective theory which is an expansion in spacetime derivatives of three scalar fields.
The condition of invariance under spacetime diffeomorphisms and internal volume preserving spatial diffeomorphisms is sufficient to characterize completely the formal structure of the theory. At the lowest order in derivatives, the action for perfect fluids is given by just an arbitrary function $F$ of a single operator that takes the form of a determinant because of  symmetry requirements. 
Identifying the various thermodynamical quantities, it turns out that $F$ is minus the energy density and is a function of the entropy density.  Its first and second derivatives are enough to characterize the pressure,  the temperature and the speed of sound of the fluid. Higher order derivatives enter in the self-interactions of the perturbations. Different types of fluids can thus be described by adequate choices of the  functional dependence of the energy density on the entropy density.  Standard cold dark matter and radiation, just to mention two simple important examples, are easily described by specific power functions.  

Considering the high precision of ongoing (e.g. \cite{Ade:2011ah}) and upcoming (e.g \cite{Laureijs:2011mu}) observations, the need for an accurate theoretical understanding of the Universe's content is more important than ever. In order to interpret the data appropriately, a framework that describes the cosmological dynamics of the matter (energy-momentum tensor) in wide generality can be a leap forward. A natural and widely popular approach has been that of using general fluids (perfect or not) \cite{Hu:1998kj} in a phenomenological and quite ad-hoc fashion to fit different types of data. As an important example, many of the studies that try to determine the properties of dark energy are based on the choice of a fluid with a (possibly time-varying) equation of state close to $-1$ and some model dependent assumptions for its perturbation properties.
Conversely, the flexibility of the formalism presented here can be relevant for model comparison and parameter estimation since the physical properties of a perfect fluid are encoded in a single function that defines the action. Instead of the common phenomenological approach of fitting several parameters like relative densities, equations of state and speed of sounds of an ensemble of fluids, one could directly work with the functions $F$'s that define the fluids. Therefore, we think that the framework presented in this paper can provide a first step towards a different useful approach in which the search is focused on functions $F$ that fully describe the energy momentum tensor. 

 The interaction terms between matter (at any order) and metric (at first order) perturbations can be easily extracted from the gravitational energy-momentum tensor. This is the case for scalar modes but also for vector and tensor ones, see equations (\ref{mixact}) and (\ref{allperturbations}). The knowledge of the function $F$ and its first two derivatives is enough to describe perfect fluids at first order in perturbations around the background solutions. Higher derivatives of $F$ will appear at the next orders in matter perturbations, which can be easily incorporated  by using  the simple structure of (\ref{mixact}) and (\ref{allperturbations}).
 
 The longitudinal (compressional) modes are found to be intimately related to the adiabatic modes at early times, $\zeta=\nabla\cdot\pi/3$, while the Goldstone boson $\hat\pi$ of the associated time shifts contains a piece that depends on the metric perturbations, $\hat\pi=-(\zeta+\phi)/\ch$.  We have shown that the equations that govern the evolution of adiabatic modes \eq{firsteq}-\eq{zetafinal} remain true even for non-perfect fluids which have vanishing non-adiabatic pressure (such as neutrinos at early times) provided that at least one fluid species, e.g. photons, can be described by a perfect fluid.
Transverse modes (vortices) mix with the vector metric perturbations and evolve accordingly to the dilution of vorticity under Hubble expansion.

There are several directions that are worth exploring in further depth. Among these, we find particularly interesting the possibility of adding more symmetry such that the system is more constrained and predictive. Conversely, demanding less restrictive symmetries would allow to draw more general conclusions.
Below we discuss two such examples that are relevant for almost de Sitter expansions.

\subsubsection*{More symmetry: scale invariance}
\label{moresym}

The condition of almost de~Sitter expansion, either at early times (inflation) or at late times (dark energy), is $w\simeq -1$, that is  $\bar{b}\bar{F}_b/\bar{F}\ll1$, see equation (\ref{wminus1}). It corresponds to the condition that the action is weakly depending on $b$, suggesting the idea of adding extra (approximate) symmetries that could suppress $F(b)$-type of actions\footnote{However it is important to stress that if the would-be leading term $F$ in the derivative expansion is suppressed,  then next order terms in the derivative expansion should generically be taken into account.}.
An example of such a symmetry is provided by internal scale transformations
\begin{equation}
\label{scaleinvariance}
\Phi^a\rightarrow e^{\lambda}\Phi^a\,,\qquad b\rightarrow e^{3\lambda} b\,.
\end{equation}
Notice that the choice $F(b)=\kappa \ln b$ gives an action that is invariant only up to shifting the cosmological constant, $F(b)\rightarrow F(b)+3\kappa\lambda$. The dynamics associated with such an action would thus be  symmetric under  (\ref{scaleinvariance}) only if $\kappa\ll1$ or if gravity is switched-off.

Other symmetries may also  be invoked to sort the desired Lagrangian $F(b)$ among all possible choices. 
An interesting case is e.g. Weyl symmetry
\begin{equation}
\label{weyl}
g_{\mu\nu}(x)\rightarrow e^{2\omega}g_{\mu\nu}(x)\qquad \Phi^a(x) \rightarrow \Phi^a(x)
\end{equation}
where $\omega$ is an arbitrary function of the spacetime coordinates. $\Phi$ has vanishing scaling dimension, in agreement with the fact that the identically conserved entropy current must have scaling dimension $3$.
Scale transformations (and the full conformal symmetry) is a subgroup of Weyl~$\times$~Diffeomorphisms.
Under Weyl transformations $b\rightarrow e^{-3\omega}b$ and thus the action is invariant when $F(b)\propto b^{4/3}$ which corresponds to the equation of state $p=\frac{1}{3}\rho$ of radiation. Such a fluid has indeed vanishing $T_\mu^\mu$ from the conservation of dilation current.
Notice that the diagonal (internal and spacetime) scale transformation $\lambda=\omega\Delta=$constant 
\begin{equation}
\label{diagonalscale}
g_{\mu\nu}(x)\rightarrow e^{2\omega}g_{\mu\nu}(x)\qquad \Phi^a(x) \rightarrow e^{\omega\Delta}\Phi^a(x)
\end{equation}
that gives $b\rightarrow e^{-3(1-\Delta)\omega}b$ leaves the action invariant when $\Delta\neq1$ and $F(b)$ is a homogeneous function of rank  $4/[3(1-\Delta)]$. Radiation and matter are selected requiring invariance under (\ref{diagonalscale}) with $\Delta=0$ and $\Delta=-1/3$ respectively.

 \subsubsection*{Less symmetry: solids}
It is interesting to consider the possibility of reducing the symmetry of the system by demoting the volume-preserving internal diffeomorphisms (\ref{diffs}) to be an approximate symmetry. This would correspond to a homogeneous and isotropic solid, i.e. a jelly, with very small transverse speed of sound $c_T^2\ll 1$.
For a jelly the dynamics is still invariant under translations (\ref{homogeneity}) and rotations (\ref{isotropy}) and the action  depends not only on $b$ but also on the traces of $B$ and $B^2$
\begin{equation}
\mathcal{L}=F(b,\mbox{Tr }B, \mbox{Tr }B^2)\,.
\end{equation}
 Assuming e.g. $\mathcal{L}=F(b)+\epsilon\mbox{Tr} B$ one generates a $\mathcal{O}(\epsilon)$ speed of sound for the transverse modes $\pi_\perp$.  This could be useful within inflationary models when trying to quantize the perfect fluid since it allows to damp the wild quantum transverse fluctuations around the classical vacuum $\Phi^i=x^i$ \cite{Endlich:2010hf}.
 
\subsubsection*{Note added} After completion of this manuscript,  a related work treating the effective theory of solids for inflation appeared \cite{solidinflation}.

\section*{Acknowledgments}
We thank Ruth Durrer, Leonardo Senatore and Filippo Vernizzi  for useful comments/suggestions; Claudia Hagedorn, Massimo Pietroni and Antonio Riotto for reading a preliminary version of the text. We also thank Lorenzo Mercolli for discussions. 
 B.B. is supported in part by the ERC Advanced Grant no.~267985, ``Electroweak Symmetry Breaking, Flavour and Dark Matter: One Solution for Three Mysteries'' (DaMeSyFla). B.B. thanks the Aspen Center for Physics for warm hospitality where part of this work was completed. The Aspen Center for Physics is supported by the National Science Foundation Grant No. 1066293.

\appendix 

\section*{Conventions and notation}
We use natural units so the speed of light is simply $c=1$ and the reduced Planck constant is $\hbar=h/(2\pi)=1$. Our choice of  Lorentzian signature is mostly positive: $(-,+,+,+)$\,. Tensor components in 4-dimensional spacetime are denoted by Greek indices running from 0 to 3, with 0 corresponding to the time coordinate. Latin indices, running from 1 to 3, are used for the purely spatial part of a 4-dimensional tensor or quantities intrinsically defined in $\mathbb{R}^3$ (such as $T_{ijk\ldots}$) or in the internal fluid space $\mathcal{F}$ (e.g. $T_{abc\ldots}$).  In any case, Greek or Latin, repeated indices in a product will be summed over by convention, unless it is otherwise specified. The symbol $\nabla$ denotes the 3-dimensional spatial gradient. Instead, 4-dimensional covariant derivatives are denoted with a semi-colon subscript. For example, the covariant derivative of a 4-vector $n^\mu$ is denoted by $n^\mu_{\,\,\,;\,\mu}$\,. We  use  $\epsilon_{\mu\nu\rho\sigma}$ to define the 4-dimensional totally antisymmetric Levi-Civita symbol with $\epsilon_{0123}=1$. The symbol  $\epsilon^{\mu\nu\rho\sigma}$ is totally antisymmetric and has $\epsilon^{0123}=-1$. The 3-dimensional totally antisymmetric Levi-Civita symbol is $\epsilon_{ijk}\equiv \epsilon_{0ijk}$.

We distinguish a background quantity form its full inhomogeneous counterpart by denoting the first one with an overbar. For instance, $\bar\rho$ denotes a background (homogeneous) density while $\rho$ denotes the corresponding inhomogeneous variable.


Partial derivatives with respect to conformal time $\tau$ are denoted with overdots. For example $\ch\equiv \dot a/a$ is the conformal Hubble parameter. 

\section{Relativistic hydrodynamics and the energy-momentum tensor}
\label{appa}

We now interpret the fluid defined in Section \ref{effective}, equation \eq{lagrangian},  in the context of  standard general relativistic hydrodynamics.  The scope of this section is  understanding how restrictive are our assumptions of homogeneity, isotropy and invariance under volume preserving diffeomorphisms of the internal $\Phi$ coordinates. 
As it is well known, see e.g. \cite{Bruni:1992dg}, any energy-momentum tensor  can be expressed in the following covariant way
\begin{align}
T_{\mu\nu}=\rho\, u_\mu u _\nu+p\, h_{\mu\nu}+u_{\mu}\,q_{\nu}+u_{\nu}\,q_{\mu}+\pi_{\mu\nu}
\end{align}
where $u^\mu$ is a choice of frame (a timelike four-vector)  and $h_{\mu\nu}=g_{\mu\nu}+u_\mu u_\nu$ is a projector on hypersurfaces orthogonal to $u^\mu$. The energy density, pressure, heat flux and anisotropic stress are respectively defined as the following projections of the energy momentum tensor:
\begin{align}
\rho=T_{\mu\nu} u^\mu u^\nu\qquad 3p=T_{\mu\nu} h^{\mu\nu}\qquad q_\mu=-h_\mu^\nu T_{\nu\gamma} u^\gamma\qquad\pi_{\mu\nu}=h_\mu^\alpha h_\nu^\beta T_{\alpha\beta}-\frac{1}{3}h_{\mu\nu}\,h_{\alpha\beta}T^{\alpha\beta}\,.
\end{align}
By construction the heat flux is orthogonal to the four-vector with respect to which we decompose the energy momentum tensor: $q_\mu u^\mu=0$. Assuming the weak energy condition
\begin{align}
T_{\mu\nu}U^\mu U^\nu\geq 0\qquad U^\mu U_\mu=-1 \qquad U^0\geq 0
\end{align}
 (for all future pointing timelike four-vectors $U^\mu$), there is a unique timelike eigenvector $u_{(E)}^\mu$ of $T_{\mu\nu}$ that is of unit norm, $u_{(E)}^2=-1$. This four-vector defines the so called energy (or rest) frame of the fluid \cite{landau1987course}. In this frame the heat flux is zero, $q_{(E)}^\mu=0$. For a perfect fluid (or one that is in thermodynamical equilibrium) the particle flux and the entropy flux are both parallel to $u^\mu_{(E)}$ which thus defines the unique choice of hydrodynamical four-velocity for the fluid\footnote{If the entropy or particle flows pointed in other directions, it would be possible to define other frames, following the world lines of these flows.} \cite{Andersson:2006nr,Bruni:1992dg,Comer:1993sp}. This can be expressed as
\begin{align} \label{44vel}
u^\mu=\frac{d x^\mu}{d \eta}
\end{align}
where $d\eta=\sqrt{-ds^2}$ is the proper time. The energy momentum tensor of a perfect fluid (i.e. with $\pi_{\mu\nu}=0$) is then 
\begin{align}\label{perfect}
T_{\mu\nu}=(\rho+p)\, u_{(E)\mu} u_{(E)\nu}+p\, g_{\mu\nu}\,.
\end{align}
When dealing with perfect fluids we drop the subscript $_{(E)}$ to simplify the notation.
From \eq{perfect} we see that imperfect fluids in their own rest frame display anisotropic stress, which can originate from dissipation or non-gravitational interactions with other fluids \cite{Ellis:1989jt}. 

Perfect fluids like \eq{perfect} are barotropic if the energy density and pressure are functionally related, $p=f(\rho)$.
Non-barotropic perfect fluids still admits an equation of state but the energy density (or pressure) in the rest frame does not completely characterize the thermodynamical properties of the system.
The perfect fluid defined by \eq{lagrangian} is barotropic since both $p$ and $\rho$ are functions of the single variable $b$.

\section{From Euler to Lagrange}
\label{appEulerLagrange}

In this appendix we comment further on the relation between the Eulerian and the Lagrangian formulation of the dynamics of perfect fluids. Going from the proper time $d\eta=\sqrt{-ds^2}$ to the conformal time $\tau$, and from the comoving coordinates $\Phi$ to the physical space coordinates $x$, by means of equation (\ref{eqforb}): 
\begin{equation}
\sqrt{-g}\, d^4 x=\frac{d^3\Phi}{b} d\eta
\end{equation}
in the FLRW background metric. Actually, this result is valid for arbitrary metrics, as it is clear from (\ref{usADM}) and (\ref{eulertolagrangemetric}) in Appendix~\ref{vorticitymetric}.
The action (\ref{action}) can thus be rewritten in a Lagrangian formulation
\begin{equation}
\label{lagrangeS}
S_{m}=\int  d^4 x\,\sqrt{-g}\, \mathcal{L}_m=
-\int d^3\Phi\, d\eta\, \frac{\rho}{b}\,,
\end{equation}
where we have used equation (\ref{rho}), which tells us that the Lagrangian $\mathcal{L}_m=F$ is equal to the comoving energy density $\rho=T_{\mu\nu}u^\mu u^\nu$ multiplid by $-1$.
Notice that $d^3\Phi/b$ is the invariant 3-volume form on spatial sections at constant times. This means that
\begin{equation}
\frac{d^3\Phi}{b(\Phi)}=\frac{d^3 f}{b(f)}\,,
\end{equation}
where $\Phi^a\rightarrow  f^a(\Phi)$ is a 3-dimensional diffeomorphism.

We see from equation (\ref{lagrangeS}) that the action  for the fluid is just the continuum limit of a ($N$-body) ``mechanical'' system made of point-like particles carrying no charges and being characterized only by their trajectories in spacetime
\begin{equation}
S_N=-\sum_{i=1}^{N}m_i \int d\eta_i \longrightarrow S_{m}=-\int d\eta\frac{d^3\Phi}{b}\rho\,.
\end{equation}
Clearly, the continuum limit has the advantage, over the corresponding discrete (and relativistic) $N$-body problem, that gravitational interactions and the matter backreaction are easily taken into account by promoting the metric to a dynamical variable and performing a local perturbation theory as presented in Section~\ref{metricp}. Moreover, while $N$-body problems deal with $3N$ 1-dimensional degrees of freedom, the continuum limit describes the system in terms of just 3 degrees of freedom in a 4-dimensional manifold. The conservation of vorticity renders trivial the dynamics of 2 degrees of freedom.

\section{Vorticity conservation}
\label{vorticitymetric}

In this appendix we use the ADM formalism \cite{Arnowitt:1962hi} to provide the expressions (valid for any metric) of the conserved charges associated with the internal volume-preserving diffeomorphisms (\ref{diffs}). 

In the ADM parameterization, the metric reads
\begin{equation}
ds^2=-N^2 d\tau^2+(dx^i+N^i d\tau)h_{ij} (dx^j+N^j d\tau)
\end{equation}
and its inverse has components
\begin{equation}
g^{00}=-\frac{1}{N^2}\,,\qquad g^{i0}=\frac{N^i}{N^2}\,,\qquad g^{ij}=h^{ij}-\frac{N^i N^j}{N^2}\,.
\end{equation}
The determinant of the metric is given by $\sqrt{-g}=N\sqrt{\det h}$.
We define the three-dimensional vector 
\begin{align}
\xi^i=\frac{1}{N}(\dot{x}^i+N^i)\,,
\end{align}
in terms of which the fluid four-velocity $u^\mu=d x^\mu/d \eta$ (being $\eta$ conformal time) is
\begin{align}
\label{usADM}
u^0=\frac{1}{N\sqrt{1-\xi^2}}\,, \qquad u^i= \frac{\dot{x}^i}{N\sqrt{1-\xi^2}}
\end{align}
where
\begin{align}
 \qquad \dot{x}^i=-[(\partial\Phi^T)^{-1}]^{i}_a\dot\Phi^a=-\frac{\partial x^i}{\partial\Phi^a}\dot\Phi^a
\end{align}
and, as in Section~\ref{flrw}, $\partial\Phi$ is the matrix of elements $(\partial\Phi)^{\,\,a}_i=\partial_i\Phi^a$.
 The scalar product $\xi^2$ is computed according to the induced spatial metric $h$,
\begin{align}
\xi^2=\xi^i \xi^j h_{ij}\,.
\end{align}

This allows to write concise expressions for the matrix $B$ defined by \eq{bmatrix}, its inverse $B^{-1}$ and the square root of its determinant \eq{smallb}:
\begin{align} \label{lb1}
B &=\partial\Phi^T\left[h^{-1}-\xi \otimes \xi \right]\partial\Phi\,,\\\label{lb2}
B^{-1} &= (\partial\Phi)^{-1}\left[h+u\otimes u\right](\partial\Phi^T)^{-1}\,,\\
b &= \det(\partial\Phi)\frac{\sqrt{1-\xi^2}}{\sqrt{\det h}}\,. \label{lb3}
\end{align}
The outer products that appear in the expressions \eq{lb1} and \eq{lb2} are $[\xi\otimes \xi]^{ij}=\xi^i \xi^j$ and  $[u\otimes u]_{ij}=u_i u_j$ with $u_{i}=g_{i\mu}u^\mu=h_{ij}\xi^j/\sqrt{1-\xi^2}$. The inverse spatial metric, that we denote $h^{-1}$, has elements $h^{ij}$ (such that $h^{ij}h_{jk}=\delta^i_k$). The spatial indexes of purely three-dimensional vectors are lowered by the spatial metric, $\xi_i=h_{ij}\xi^j$ and $N_i=h_{ij}N^j=g_{0i}$.

As discussed in Section~\ref{effective}, any  function $\varepsilon^a(\Phi)$ gives rise to a current
\begin{equation}
\mathcal{J}^\mu_{(\varepsilon)}=-b F_b\, \varepsilon^c(\Phi) (B^{-1})^{cd} \partial^\mu\Phi^d 
\end{equation}
which is conserved on the equations of motion (\ref{eom}) as long as $\partial_a\varepsilon^a=0$.
The transversality constraint (\ref{epsilondiff}) ensures that $\Phi^a\rightarrow \Phi^a+\varepsilon^a(\Phi)$ is, locally, a volume-preserving diffeomorphism.
The associated conserved charges are defined in the usual way
\begin{equation}
Q_{(\varepsilon)}=-\int d^3x\sqrt{\det h} \,\mathcal{J}^\mu_{(\varepsilon)} n_\mu\qquad \dot{Q}_{(\varepsilon)}=0\,,
\end{equation}
where $n^\mu=1/N(1,-N^{i})$ is the normal unit vector to the constant time hypersurface $\Sigma_\tau$ that has $h$ as the induced spatial metric. 

Using \eq{lb3} to pass to the Lagrangian formulation:
\begin{equation}
\label{eulertolagrangemetric}
d^3 x= \frac{d^3\Phi}{b}\frac{\sqrt{1-\xi^2}}{\sqrt{\det h}}\,,
\end{equation}
the conserved charges can be expressed in a simple form
\begin{equation}
Q_{(\varepsilon)}=-\int \frac{d^3\Phi}{b} \sqrt{1-\xi^2}\frac{1}{N}\left(\mathcal{J}_0-N^i\mathcal{J}_i\right)=
\int d^3\Phi \frac{(\rho+p)}{b} \varepsilon^{a}\frac{\partial x^i}{\partial \Phi^a}u_i\,.
\end{equation}
Solving the constraint (\ref{epsilondiff}) with 
$\varepsilon^a(\Phi)=\epsilon_{abc}\partial_b  f_{c}(\Phi)$ the charges take the following form after integrating by parts:
\begin{equation}
\label{generalcharge}
Q_{(f)}=
\int d^3\Phi \,\,f_{a}(\Phi) \epsilon_{abc}  \partial_b\left[ \frac{(\rho+p)}{b} \frac{\partial x^i}{\partial \Phi^c}u_i \right]  \,.
\end{equation}

Let us now move on to the specific case of the vorticity currents, generalizing the results of Section~\ref{flrw}. 
The vorticity charges are defined by picking a delta function in (\ref{generalcharge}), $f_a=\alpha_a \delta^3(\Phi-\tilde\Phi)$. The resulting vorticity charges are given by the \textit{curl}, in the three-dimensional $\Phi$-space, of  a (co)vector $V$
\begin{equation}
Q_a=\epsilon_{abc}\partial_b V_c\,, \qquad V_a=\frac{(\rho+p)}{b} \frac{\partial x^i}{\partial \Phi^a}u_i\,,\qquad \dot{Q}_a=0\,,
\end{equation}
where we have renamed $\tilde\Phi\rightarrow \Phi$. By comparing these expressions with (\ref{generalcharge}), we see that any conserved charge can be expressed  as a linear functional of the vorticity charges $Q_a$:
\begin{equation}
Q_{(f)}=
\int d^3\Phi f_{a}(\Phi) Q_a(\Phi)\,.
\end{equation}
Notice that the leading order term in perturbations for $Q$, around the metric (\ref{unpfrw}) with the (background) field $\bar{\Phi}^a=x^a$ choice, gives
\begin{equation}
\label{Qperturb}
Q_a=-\left[ a^4(\bar\rho+\bar{p})\epsilon_{aij}\partial_i \left(\dot\pi_{\perp}^j-\frac{\delta g_{0j}}{a^2}\right)\right]+\ldots
\end{equation}
 where the ellipsis contain all the higher order terms. This shows explicitly that in the Poisson gauge  (\ref{poissongauge}), the vorticity conservation is precisely equivalent to the dynamical equation (\ref{vecttrans}) for the transverse modes.

The common definition of vorticity circulation $\mathcal{V}$ can now be naturally extended to arbitrary spacetime metrics: it is the  \textit{flux} of the vorticity $Q$ over a  surface $\mathcal{S}$ (with boundary $\partial\mathcal{S}$) in the internal $\Phi$-space   
\begin{align}
\label{vorticitydefmetric}
\mathcal{V}  \equiv  \int_{\mathcal{S}} \star (Q_a d\Phi^a)=\oint_{\partial\mathcal{S}} V_a d\Phi^a=\oint_{\Gamma}\frac{(\rho+p)}{b} u_i dx^i
\end{align}
where the boundary $\partial\mathcal{S}$ is mapped into a closed loop, $\Gamma=x(\tau,\partial\mathcal{S})$,  in real space. 
Since $Q$ is constant on-shell, the circulation $\mathcal{V}$ is conserved on the solutions of the equations of motion.
This shows that the results of \cite{Dubovsky:2005xd} for Minkowski spacetime carry over arbitrary metrics that admit a 3+1 decomposition.


\end{document}